\documentstyle[12pt,aasms4]{article}
\input epsf.tex
\def\temp{1.35}%
\let\tempp=\relax
\expandafter\ifx\csname psboxversion\endcsname\relax
\else
    \ifdim\temp cm>\psboxversion cm
    \else
      \let\temp=\psboxversion
      \let\tempp= 
    \fi
\fi
\tempp
\let\psboxversion=\temp
\catcode`\@=11
\def\psfortextures{
\def\PSspeci@l##1##2{%
\special{illustration ##1\space scaled ##2}%
}}%
\def\psfordvitops{
\def\PSspeci@l##1##2{%
\special{dvitops: import ##1\space \the\drawingwd \the\drawinght}%
}}%
\def\psfordvips{
\def\PSspeci@l##1##2{%
\d@my=0.1bp \d@mx=\drawingwd \divide\d@mx by\d@my
\includegraphics{##1\space}}}%
\def\psforoztex{
\def\PSspeci@l##1##2{%
\special{##1 \space
      ##2 1000 div dup scale
      \number-\psllx\space\space \number-\pslly\space\space translate
}}}%
\def\psfordvitps{
\def\dvitpsLiter@ldim##1{\dimen0=##1\relax
\special{dvitps: Literal "\number\dimen0\space"}}%
\def\PSspeci@l##1##2{%
\at(0bp;\drawinght){%
\special{dvitps: Include0 "psfig.psr"}
\dvitpsLiter@ldim{\drawingwd}%
\dvitpsLiter@ldim{\drawinght}%
\dvitpsLiter@ldim{\psllx bp}%
\dvitpsLiter@ldim{\pslly bp}%
\dvitpsLiter@ldim{\psurx bp}%
\dvitpsLiter@ldim{\psury bp}%
\special{dvitps: Literal "startTexFig"}%
\special{dvitps: Include1 "##1"}%
\special{dvitps: Literal "endTexFig"}%
}}}%
\def\psfordvialw{
\def\PSspeci@l##1##2{
\special{language "PostScript",
position = "bottom left",
literal "  \psllx\space \pslly\space translate
  ##2 1000 div dup scale
  -\psllx\space -\pslly\space translate",
include "##1"}
}}%
\def\psforptips{
\def\PSspeci@l##1##2{{
\d@mx=\psurx bp
\advance \d@mx by -\psllx bp
\divide \d@mx by 1000\multiply\d@mx by \xscale
\incm{\d@mx}
\let\tmpx\dimincm
\d@my=\psury bp
\advance \d@my by -\pslly bp
\divide \d@my by 1000\multiply\d@my by \xscale
\incm{\d@my}
\let\tmpy\dimincm
\d@mx=-\psllx bp
\divide \d@mx by 1000\multiply\d@mx by \xscale
\d@my=-\pslly bp
\divide \d@my by 1000\multiply\d@my by \xscale
\at(\d@mx;\d@my){\special{ps:##1 x=\tmpx cm, y=\tmpy cm}}
}}}%
\def\psonlyboxes{
\def\PSspeci@l##1##2{%
\at(0cm;0cm){\boxit{\vbox to\drawinght
  {\vss\hbox to\drawingwd{\at(0cm;0cm){\hbox{({\tt##1})}}\hss}}}}
}}%
\def\psloc@lerr#1{%
\let\savedPSspeci@l=\PSspeci@l%
\def\PSspeci@l##1##2{%
\at(0cm;0cm){\boxit{\vbox to\drawinght
  {\vss\hbox to\drawingwd{\at(0cm;0cm){\hbox{({\tt##1}) #1}}\hss}}}}
\let\PSspeci@l=\savedPSspeci@l
}}%
\newread\pst@mpin
\newdimen\drawinght\newdimen\drawingwd
\newdimen\psxoffset\newdimen\psyoffset
\newbox\drawingBox
\newcount\xscale \newcount\yscale \newdimen\pscm\pscm=1cm
\newdimen\d@mx \newdimen\d@my
\newdimen\pswdincr \newdimen\pshtincr
\let\ps@nnotation=\relax
{\catcode`\|=0 |catcode`|\=12 |catcode`|
|catcode`#=12 |catcode`*=14
|xdef|backslashother{\}*
|xdef|percentother{
|xdef|tildeother{~}*
|xdef|sharpother{#}*
}%
\def\R@moveMeaningHeader#1:->{}%
\def\uncatcode#1{%
\edef#1{\expandafter\R@moveMeaningHeader\meaning#1}}%
\def\execute#1{#1}
\def\psm@keother#1{\catcode`#112\relax}
\def\executeinspecs#1{%
\execute{\begingroup\let\do\psm@keother\dospecials\catcode`\^^M=9#1\endgroup}}%
\def\@mpty{}%
\def\matchexpin#1#2{
  \fi%
  \edef\tmpb{{#2}}%
  \expandafter\makem@tchtmp\tmpb%
  \edef\tmpa{#1}\edef\tmpb{#2}%
  \expandafter\expandafter\expandafter\m@tchtmp\expandafter\tmpa\tmpb\endm@tch%
  \if\match%
}%
\def\matchin#1#2{%
  \fi%
  \makem@tchtmp{#2}%
  \m@tchtmp#1#2\endm@tch%
  \if\match%
}%
\def\makem@tchtmp#1{\def\m@tchtmp##1#1##2\endm@tch{%
  \def\tmpa{##1}\def\tmpb{##2}\let\m@tchtmp=\relax%
  \ifx\tmpb\@mpty\def\match{YN}%
  \else\def\match{YY}\fi%
}}%
\def\incm#1{{\psxoffset=1cm\d@my=#1
 \d@mx=\d@my
  \divide\d@mx by \psxoffset
  \xdef\dimincm{\number\d@mx.}
  \advance\d@my by -\number\d@mx cm
  \multiply\d@my by 100
 \d@mx=\d@my
  \divide\d@mx by \psxoffset
  \edef\dimincm{\dimincm\number\d@mx}
  \advance\d@my by -\number\d@mx cm
  \multiply\d@my by 100
 \d@mx=\d@my
  \divide\d@mx by \psxoffset
  \xdef\dimincm{\dimincm\number\d@mx}
}}%
\newif\ifNotB@undingBox
\newhelp\PShelp{Proceed: you'll have a 5cm square blank box instead of
your graphics.}%
\def\s@tsize#1 #2 #3 #4\@ndsize{
  \def\psllx{#1}\def\pslly{#2}%
  \def\psurx{#3}\def\psury{#4}
  \ifx\psurx\@mpty\NotB@undingBoxtrue
  \else
    \drawinght=#4bp\advance\drawinght by-#2bp
    \drawingwd=#3bp\advance\drawingwd by-#1bp
  \fi
  }%
\def\sc@nBBline#1:#2\@ndBBline{\edef\p@rameter{#1}\edef\v@lue{#2}}%
\def\g@bblefirstblank#1#2:{\ifx#1 \else#1\fi#2}%
{\catcode`\%=12
\xdef\B@undingBox{
\def\ReadPSize#1{
 \readfilename#1\relax
 \let\PSfilename=\lastreadfilename
 \openin\pst@mpin=#1\relax
 \ifeof\pst@mpin \errhelp=\PShelp
   \errmessage{I haven't found your postscript file (\PSfilename)}%
   \psloc@lerr{was not found}%
   \s@tsize 0 0 142 142\@ndsize
   \closein\pst@mpin
 \else
   \if\matchexpin{\GlobalInputList}{, \lastreadfilename}%
   \else\xdef\GlobalInputList{\GlobalInputList, \lastreadfilename}%
     \immediate\write\psbj@inaux{\lastreadfilename,}%
   \fi%
   \loop
     \executeinspecs{\catcode`\ =10\global\read\pst@mpin to\n@xtline}%
     \ifeof\pst@mpin
       \errhelp=\PShelp
       \errmessage{(\PSfilename) is not an Encapsulated PostScript File:
           I could not find any \B@undingBox: line.}%
       \edef\v@lue{0 0 142 142:}%
       \psloc@lerr{is not an EPSFile}%
       \NotB@undingBoxfalse
     \else
       \expandafter\sc@nBBline\n@xtline:\@ndBBline
       \ifx\p@rameter\B@undingBox\NotB@undingBoxfalse
         \edef\t@mp{%
           \expandafter\g@bblefirstblank\v@lue\space\space\space}%
         \expandafter\s@tsize\t@mp\@ndsize
       \else\NotB@undingBoxtrue
       \fi
     \fi
   \ifNotB@undingBox\repeat
   \closein\pst@mpin
 \fi
\message{#1}%
}%
\def\psboxto(#1;#2)#3{\vbox{%
   \ReadPSize{#3}%
   \advance\pswdincr by \drawingwd
   \advance\pshtincr by \drawinght
   \divide\pswdincr by 1000
   \divide\pshtincr by 1000
   \d@mx=#1
   \ifdim\d@mx=0pt\xscale=1000
         \else \xscale=\d@mx \divide \xscale by \pswdincr\fi
   \d@my=#2
   \ifdim\d@my=0pt\yscale=1000
         \else \yscale=\d@my \divide \yscale by \pshtincr\fi
   \ifnum\yscale=1000
         \else\ifnum\xscale=1000\xscale=\yscale
                    \else\ifnum\yscale<\xscale\xscale=\yscale\fi
              \fi
   \fi
   \divide\drawingwd by1000 \multiply\drawingwd by\xscale
   \divide\drawinght by1000 \multiply\drawinght by\xscale
   \divide\psxoffset by1000 \multiply\psxoffset by\xscale
   \divide\psyoffset by1000 \multiply\psyoffset by\xscale
   \global\divide\pscm by 1000
   \global\multiply\pscm by\xscale
   \multiply\pswdincr by\xscale \multiply\pshtincr by\xscale
   \ifdim\d@mx=0pt\d@mx=\pswdincr\fi
   \ifdim\d@my=0pt\d@my=\pshtincr\fi
   \message{scaled \the\xscale}%
 \hbox to\d@mx{\hss\vbox to\d@my{\vss
   \global\setbox\drawingBox=\hbox to 0pt{\kern\psxoffset\vbox to 0pt{%
      \kern-\psyoffset
      \PSspeci@l{\PSfilename}{\the\xscale}%
      \vss}\hss\ps@nnotation}%
   \global\wd\drawingBox=\the\pswdincr
   \global\ht\drawingBox=\the\pshtincr
   \global\drawingwd=\pswdincr
   \global\drawinght=\pshtincr
   \baselineskip=0pt
   \copy\drawingBox
 \vss}\hss}%
  \global\psxoffset=0pt
  \global\psyoffset=0pt
  \global\pswdincr=0pt
  \global\pshtincr=0pt 
  \global\pscm=1cm 
}}%
\def\psboxscaled#1#2{\vbox{%
  \ReadPSize{#2}%
  \xscale=#1
  \message{scaled \the\xscale}%
  \divide\pswdincr by 1000 \multiply\pswdincr by \xscale
  \divide\pshtincr by 1000 \multiply\pshtincr by \xscale
  \divide\psxoffset by1000 \multiply\psxoffset by\xscale
  \divide\psyoffset by1000 \multiply\psyoffset by\xscale
  \divide\drawingwd by1000 \multiply\drawingwd by\xscale
  \divide\drawinght by1000 \multiply\drawinght by\xscale
  \global\divide\pscm by 1000
  \global\multiply\pscm by\xscale
  \global\setbox\drawingBox=\hbox to 0pt{\kern\psxoffset\vbox to 0pt{%
     \kern-\psyoffset
     \PSspeci@l{\PSfilename}{\the\xscale}%
     \vss}\hss\ps@nnotation}%
  \advance\pswdincr by \drawingwd
  \advance\pshtincr by \drawinght
  \global\wd\drawingBox=\the\pswdincr
  \global\ht\drawingBox=\the\pshtincr
  \global\drawingwd=\pswdincr
  \global\drawinght=\pshtincr
  \baselineskip=0pt
  \copy\drawingBox
  \global\psxoffset=0pt
  \global\psyoffset=0pt
  \global\pswdincr=0pt
  \global\pshtincr=0pt 
  \global\pscm=1cm
}}%
\def\psbox#1{\psboxscaled{1000}{#1}}%
\newif\ifn@teof\n@teoftrue
\newif\ifc@ntrolline
\newif\ifmatch
\newread\j@insplitin
\newwrite\j@insplitout
\newwrite\psbj@inaux
\immediate\openout\psbj@inaux=psbjoin.aux
\immediate\write\psbj@inaux{\string\joinfiles}%
\immediate\write\psbj@inaux{\jobname,}%
\def\toother#1{\ifcat\relax#1\else\expandafter%
  \toother@ux\meaning#1\endtoother@ux\fi}%
\def\toother@ux#1 #2#3\endtoother@ux{\def\tmp{#3}%
  \ifx\tmp\@mpty\def\tmp{#2}\let\next=\relax%
  \else\def\next{\toother@ux#2#3\endtoother@ux}\fi%
\next}%
\let\readfilenamehook=\relax
\def\re@d{\expandafter\re@daux}
\def\re@daux{\futurelet\nextchar\stopre@dtest}%
\def\re@dnext{\xdef\lastreadfilename{\lastreadfilename\nextchar}%
  \afterassignment\re@d\let\nextchar}%
\def\stopre@d{\egroup\readfilenamehook}%
\def\stopre@dtest{%
  \ifcat\nextchar\relax\let\nextread\stopre@d
  \else
    \ifcat\nextchar\space\def\nextread{%
      \afterassignment\stopre@d\chardef\nextchar=`}%
    \else\let\nextread=\re@dnext
      \toother\nextchar
      \edef\nextchar{\tmp}%
    \fi
  \fi\nextread}%
\def\readfilename{\bgroup%
  \let\\=\backslashother \let\%=\percentother \let\~=\tildeother
  \let\#=\sharpother \xdef\lastreadfilename{}%
  \re@d}%
\xdef\GlobalInputList{\jobname}%
\def\psnewinput{%
  \def\readfilenamehook{
    \if\matchexpin{\GlobalInputList}{, \lastreadfilename}%
    \else\xdef\GlobalInputList{\GlobalInputList, \lastreadfilename}%
      \immediate\write\psbj@inaux{\lastreadfilename,}%
    \fi%
    \let\readfilenamehook=\relax%
    \ps@ldinput\lastreadfilename\relax%
  }\readfilename%
}%
\expandafter\ifx\csname @@input\endcsname\relax    
  \immediate\let\ps@ldinput=\input\def\input{\psnewinput}%
\else
  \immediate\let\ps@ldinput=\@@input
  \def\@@input{\psnewinput}%
\fi%
\def\nowarnopenout{%
 \def\warnopenout##1##2{%
   \readfilename##2\relax
   \message{\lastreadfilename}%
   \immediate\openout##1=\lastreadfilename\relax}}%
\def\warnopenout#1#2{%
 \readfilename#2\relax
 \def\t@mp{TrashMe,psbjoin.aux,psbjoint.tex,}\uncatcode\t@mp
 \if\matchexpin{\t@mp}{\lastreadfilename,}%
 \else
   \immediate\openin\pst@mpin=\lastreadfilename\relax
   \ifeof\pst@mpin
     \else
     \edef\tmp{{If the content of this file is precious to you, this
is your last chance to abort (ie press x or e) and rename it before
retexing (\jobname). If you're sure there's no file
(\lastreadfilename) in the directory of (\jobname), then go on: I'm
simply worried because you have another (\lastreadfilename) in some
directory I'm looking in for inputs...}}%
     \errhelp=\tmp
     \errmessage{I may be about to replace your file named \lastreadfilename}%
   \fi
   \immediate\closein\pst@mpin
 \fi
 \message{\lastreadfilename}%
 \immediate\openout#1=\lastreadfilename\relax}%
{\catcode`\%=12\catcode`\*=14
\gdef\splitfile#1{*
 \readfilename#1\relax
 \immediate\openin\j@insplitin=\lastreadfilename\relax
 \ifeof\j@insplitin
   \message{! I couldn't find and split \lastreadfilename!}*
 \else
   \immediate\openout\j@insplitout=TrashMe
   \message{< Splitting \lastreadfilename\space into}*
   \loop
     \ifeof\j@insplitin
       \immediate\closein\j@insplitin\n@teoffalse
     \else
       \n@teoftrue
       \executeinspecs{\global\read\j@insplitin to\spl@tinline\expandafter
         \ch@ckbeginnewfile\spl@tinline
       \ifc@ntrolline
       \else
         \toks0=\expandafter{\spl@tinline}*
         \immediate\write\j@insplitout{\the\toks0}*
       \fi
     \fi
   \ifn@teof\repeat
   \immediate\closeout\j@insplitout
 \fi\message{>}*
}*
\gdef\ch@ckbeginnewfile#1
 \def\t@mp{#1}*
 \ifx\@mpty\t@mp
   \def\t@mp{#3}*
   \ifx\@mpty\t@mp
     \global\c@ntrollinefalse
   \else
     \immediate\closeout\j@insplitout
     \warnopenout\j@insplitout{#2}*
     \global\c@ntrollinetrue
   \fi
 \else
   \global\c@ntrollinefalse
 \fi}*
\gdef\joinfiles#1\into#2{*
 \message{< Joining following files into}*
 \warnopenout\j@insplitout{#2}*
 \message{:}*
 {*
 \edef\w@##1{\immediate\write\j@insplitout{##1}}*
\w@{
\w@{
\w@{
\w@{
\w@{
\w@{
\w@{
\w@{
\w@{
\w@{
\w@{\string\input\space psbox.tex}*
\w@{\string\splitfile{\string\jobname}}*
\w@{\string\let\string\autojoin=\string\relax}*
}*
 \expandafter\tre@tfilelist#1, \endtre@t
 \immediate\closeout\j@insplitout
 \message{>}*
}*
\gdef\tre@tfilelist#1, #2\endtre@t{*
 \readfilename#1\relax
 \ifx\@mpty\lastreadfilename
 \else
   \immediate\openin\j@insplitin=\lastreadfilename\relax
   \ifeof\j@insplitin
     \errmessage{I couldn't find file \lastreadfilename}*
   \else
     \message{\lastreadfilename}*
     \immediate\write\j@insplitout{
     \executeinspecs{\global\read\j@insplitin to\oldj@ininline}*
     \loop
       \ifeof\j@insplitin\immediate\closein\j@insplitin\n@teoffalse
       \else\n@teoftrue
         \executeinspecs{\global\read\j@insplitin to\j@ininline}*
         \toks0=\expandafter{\oldj@ininline}*
         \let\oldj@ininline=\j@ininline
         \immediate\write\j@insplitout{\the\toks0}*
       \fi
     \ifn@teof
     \repeat
   \immediate\closein\j@insplitin
   \fi
   \tre@tfilelist#2, \endtre@t
 \fi}*
}%
\def\autojoin{%
 \immediate\write\psbj@inaux{\string\into{psbjoint.tex}}%
 \immediate\closeout\psbj@inaux
 \expandafter\joinfiles\GlobalInputList\into{psbjoint.tex}%
}%
\def\centinsert#1{\midinsert\line{\hss#1\hss}\endinsert}%
\def\psannotate#1#2{\vbox{%
  \def\ps@nnotation{#2\global\let\ps@nnotation=\relax}#1}}%
\def\pscaption#1#2{\vbox{%
   \setbox\drawingBox=#1
   \copy\drawingBox
   \vskip\baselineskip
   \vbox{\hsize=\wd\drawingBox\setbox0=\hbox{#2}%
     \ifdim\wd0>\hsize
       \noindent\unhbox0\tolerance=5000
    \else\centerline{\box0}%
    \fi
}}}%
\def\at(#1;#2)#3{\setbox0=\hbox{#3}\ht0=0pt\dp0=0pt
  \rlap{\kern#1\vbox to0pt{\kern-#2\box0\vss}}}%
\newdimen\gridht \newdimen\gridwd
\def\gridfill(#1;#2){%
  \setbox0=\hbox to 1\pscm
  {\vrule height1\pscm width.4pt\leaders\hrule\hfill}%
  \gridht=#1
  \divide\gridht by \ht0
  \multiply\gridht by \ht0
  \gridwd=#2
  \divide\gridwd by \wd0
  \multiply\gridwd by \wd0
  \advance \gridwd by \wd0
  \vbox to \gridht{\leaders\hbox to\gridwd{\leaders\box0\hfill}\vfill}}%
\def\fillinggrid{\at(0cm;0cm){\vbox{%
  \gridfill(\drawinght;\drawingwd)}}}%
\def\textleftof#1:{%
  \setbox1=#1
  \setbox0=\vbox\bgroup
    \advance\hsize by -\wd1 \advance\hsize by -2em}%
\def\textrightof#1:{%
  \setbox0=#1
  \setbox1=\vbox\bgroup
    \advance\hsize by -\wd0 \advance\hsize by -2em}%
\def\endtext{%
  \egroup
  \hbox to \hsize{\valign{\vfil##\vfil\cr%
\box0\cr%
\noalign{\hss}\box1\cr}}}%
\def\frameit#1#2#3{\hbox{\vrule width#1\vbox{%
  \hrule height#1\vskip#2\hbox{\hskip#2\vbox{#3}\hskip#2}%
        \vskip#2\hrule height#1}\vrule width#1}}%
\def\boxit#1{\frameit{0.4pt}{0pt}{#1}}%
\catcode`\@=12 
\psfordvips   

\newcommand\nion[2]{#1\,\lowercase{{\sc #2}}}
\newcommand\wave[1]{\mbox{$\lambda$#1\,\AA}}
\newcommand\iso[2]{${\rm ^{#2}}$#1}
\newcommand\eps[1]{log~$\varepsilon$(#1)}
\def\kmsec{\mbox{km~s$^{\rm -1}$}}
\def\teff{\mbox{T$_{\rm eff}$}}
\def\vt{\mbox{v$_{\rm t}$}}
\def\BmV0{\mbox{(B-V)$^{\rm o}$}}
\def\VmK0{\mbox{(V-K)$^{\rm o}$}}
\def\MV0{\mbox{M$_{\rm V}^{\rm o}$}}
\def\MV{\mbox{M$_{\rm V}$}}
\def\Msun{\mbox{M$_{\odot}$}}
\def\carbiso{\mbox{${\rm ^{12}C/^{13}C}$}}
\def\etal{\mbox{\it{et al.}\ }}
\def\eg{\mbox{{\it e.g.}}}
\def\ie{\mbox{{\it i.e.}}}
\def\first{{\it 1$^{st}$}}
\def\second{{\it 2$^{nd}$}}
\def\third{{\it 3$^{rd}$}}
\def\oneonefive{{\rm HD~115444}}
\def\onetwotwo{{\rm HD~122563}}
\def\cs22892{{\rm CS~22892-052}}
\def\asec {^{\prime\prime}}
\def\amin {^{\prime}}
\def\fsec  {{\rlap.}^{\rm s}}
\def\fasec {{\rlap.}^{\prime \prime}\hskip0.05em}
\def\deg  {^\circ}
\def\about  {$\sim$}

\begin{document}

\title{THE FADING RADIO EMISSION FROM SN 1961V:\\
EVIDENCE FOR A TYPE II PECULIAR SUPERNOVA?}

\author{Christopher J. Stockdale\altaffilmark{1}, Michael P. Rupen\altaffilmark{2}, John J. Cowan\altaffilmark{1}, You-Hua~Chu\altaffilmark{3}, and Steven S. Jones\altaffilmark{1}
}


\altaffiltext{1}{Department of Physics and Astronomy,
440 West Brooks Room 131, University of Oklahoma, Norman, OK 73019; 
stockdal@mail.nhn.ou.edu, cowan@mail.nhn.ou.edu}

\altaffiltext{2}{National Radio Astronomical Observatories, 
PO Box 0, 1003 Lopezville Road, Socorro, NM 87801-0387; mrupen@nrao.edu}

\altaffiltext{3}{Astronomy Department, 
University of Illinois, 1002 West Green Street, Urbana, IL 61801; 
chu@astro.uiuc.edu}

\begin{abstract}

Using the Very Large Array (VLA), 
we have detected radio emission from the site of SN~1961V 
in the Sc galaxy NGC~1058.  With a peak flux density
of  $0.063\pm 0.008$ mJy/beam at 6~cm and 
$0.147\pm 0.026$ mJy/beam at 18~cm, the source is non-thermal, 
with a spectral index of $-0.79\pm 0.23$.  Within errors, this spectral
index is the 
same value reported for previous VLA observations taken in 1984 and 1986.
The radio emission at both wavelengths has decayed since the mid 1980's
observations with power-law indices of  
$\beta _{20cm} =-0.69\pm 0.23$ and $\beta _{6cm} = -1.75\pm 0.16$. 
We discuss the radio properties of this source
and compare them with those of Type II radio supernovae and luminous blue
variables. 

\end{abstract}


\section{Introduction}

Supernova (SN) 1961V, the prototype of Zwicky's Type V SNe 
(now classified as either a Type II Peculiar SN or a luminous blue variable (LBV)), was unique in several respects (\cite{bra71}).  Its progenitor
was visible as an 18th magnitude star from 1937 to 1960.  It is 
the first SN, prior to SN~1987A,
whose parent star was identified before it exploded 
(assuming a SN interpretation is correct for this event).
The bolometric correction,
the exact distance, and the extinction are all uncertain, but its pre-outburst
luminosity apparently exceeded 10$^{41}$~ergs~s$^{-1}$, which is the
Eddington limit
for a 240~M$_{\odot}$ star.  After the explosion in late 1961, the
initial peak of the optical light curve was more complex 
and much broader than for any supernova ever observed.
Subsequently, the optical light
curve decayed more slowly, by about 5 magnitudes in 8 years.   
Few SNe have been followed optically for more than 2 years. 
Optical spectra taken during this extended bright phase
showed that the characteristic expansion velocity of SN~1961V was 
2,000~km~s$^{-1}$, which differs from 
the typical value of 10,000~km~s$^{-1}$ for most SNe.  This velocity
is similar to novae expansion velocities.  However, no novae are
this strong and none have persisted for this long in the radio.  
This velocity is in fact consistent with the measurements of SN~1986J (another 
Type II Peculiar SN), which had an expansion velocity (taken well after 
maximum optical brightness) of 1,000~km~s$^{-1}$ (\cite{van86,rup87,wei88}).  

Using the Very Large Array (VLA),\footnote{The National Radio Astronomy 
Observatory is a facility of the National Science Foundation operated
under cooperative agreement by Associated Universities, Inc.}
observations of SN~1961V were made in the mid 1980s, with the most
definitive search in 1986 (\cite{cow88}) (hereafter referred to
as CHB).  CHB detected a non-thermal radio source 
at the precise position of SN~1961V.
Fesen (1985) also reported recovering SN~1961V in the optical.  
CHB later detected an optical counterpart to SN~1961V,
which was identified as an H~II region using filter photometry.
[CHB also detected another slightly fainter radio source to the west of
SN~1961V with a similar non-thermal spectral index.  This source was identified 
as a supernova remnant (SNR) not previously identified. 
This SNR also has an associated optical counterpart (i.e., an H II region).]
At the distance of NGC~1058 (9.3 Mpc) (\cite{tul80,sil96}),
SN~1961V is as radio luminous as the bright Galactic SNR Cas~A.
SN~1961V's luminosity is also comparable to several historical decades-old 
(also known as intermediate-age) radio supernovae (RSNe) including
SNe 1923A, 1950B, 1957D, 1968D, 1970G and 1986J (\cite{cow91,cow94,eck98,eck00,hym95,wei88}).

Recently, however, there has been some question about whether
the event identified as SN~1961V was actually a supernova.
Goodrich \etal (1989) suggest instead that
this event was an LBV similar to
$\eta$~Carinae,
and that the supposed supernova was an outburst of the variable
star. Subsequently, Filippenko \etal (1995) observed
SN~1961V using the Hubble Space Telescope (HST),
 although at that time the HST had not been refurbished.
Those observations seemed to suggest a (very faint) star is still
present at the site, which might or might not argue against a supernova origin.
Among the brightest LBVs (e.g. $\eta$~Car, P~Cygni, and V~12 in NGC~2403),
$\eta$~Car is reported to have been the most luminous, reaching
$M_{Bol} \simeq -14$.  In comparison, 
SN~1961V was reported to have peaked at $M_{Bol}~\simeq ~-17$
(\cite{hum94}).  This peak estimate for SN~1961V 
is likely underestimated by 1.2 magnitudes if one accounts for the 
more recently derived Cepheid distance (Silbermann \etal 1996).  
This would make
SN~1961V nearly $50\times$ brighter in the optical than $\eta$~Car at 
maximum brightness (\cite{hum94}).  

To assess the exact nature of this event we have performed a series of 
observations at various wavelengths, employing the phased-VLA with the 
Very Large Baseline 
Interferometer (VLBI) and the ROSAT X-ray satellite.
In this paper we report on our recent VLA radio observations of SN~1961V 
and what they indicate about the nature of this event.
The VLBI and ROSAT results are reported in Stockdale (2001).

\section{Observations and Results}

The new VLA data on SN~1961V are taken from three observing runs.
In the first, SN~1961V was observed for 12 hours on 14 September
1999 at 18cm (1.67 GHz) using the VLA's most extended (A) configuration, 
with a maximum baseline of 34 km.  These data were taken while
the VLA was being used in phased-array mode for a VLBI run, and the total
bandwidth was 50 MHz in each of the two orthogonal circular
polarizations.  The phase calibrator was J0253+3835, and both 3C286 and
3C48 were used to set the flux density scale.  The total time on-source 
was 4.7 hours. 

During the second pair of observing runs, on 21 and 25 January 2000, the
VLA was in its B configuration (maximum baseline of 10km), and observed
at 6cm (4.89 GHz) for a total of 12 hours.  Here we used the standard
VLA continuum mode, obtaining a total of 100 MHz bandwidth in each of
the two orthogonal circular polarizations.  The phase calibrator was
J0251+4032, and 3C48 was used to set the flux density scale.  
In all observations the pointing center was CHB's radio position for
SN~1961V, and flux densities for 3C48 and 3C286 were taken from 
Perley, Butler, \& Zijlstra (2000).

\begin{figure}
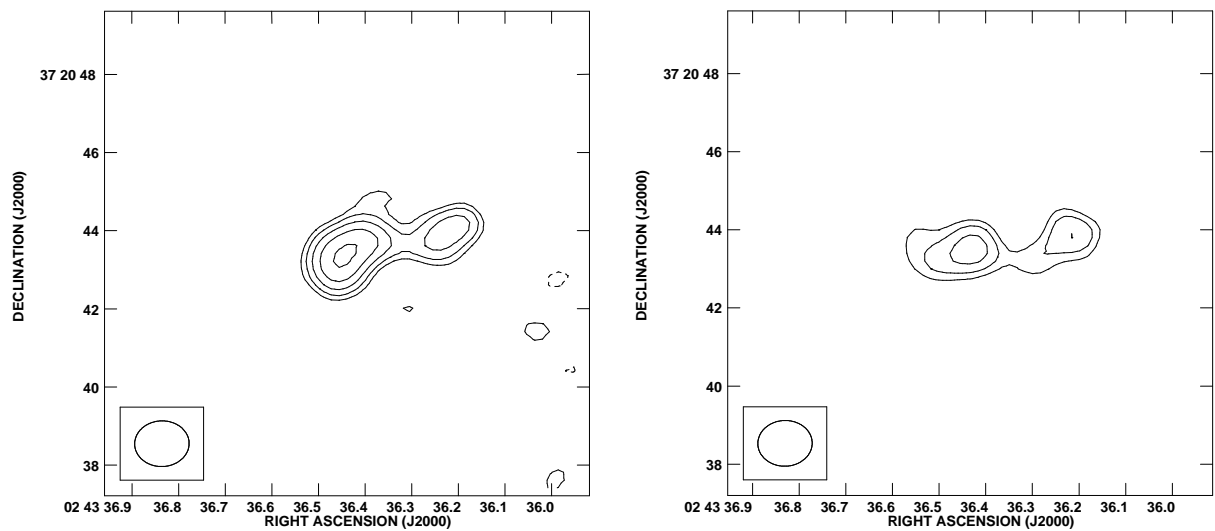

$$ \psboxto(3.1in;0in){fig1a.ps}
\quad \psboxto(0.0in;3.2in){fig1b.ps}$$

\caption{
{\it(a.)} (left) Radio contour map at 20 cm (1.5 GHz) of SN~1961V (the strong 
eastern source) and a neighboring SNR (western source).  Contour levels are -0.11, -0.08, -0.06, 0.06, 0.08, 0.11, 0.16, 0.23, and 0.32 mJy beam$^{-1}$, 
with a beam size of $1\fasec 26 \times 1\fasec 05$, p.a. = $89\deg$, and 
rms noise of 0.026 mJy beam$^{-1}$.  Observations taken with the VLA in A 
configuration 15 November 1984.
{\it(b.)} (right) Radio contour map at 18 cm (1.7 GHz) of the same region.  Contour levels are -0.11, -0.08, -0.06, 
0.06, 0.08, 0.11, 0.16, 0.23, and 0.32 mJy beam$^{-1}$, with a
beam size 
of $1\fasec 20 \times 1\fasec 01$, p.a. = $82\deg$, and rms noise of 
0.026 mJy beam$^{-1}$.  Observations 
taken with the VLA in A configuration 14 September 1999.
\label{fig1}}
\end{figure}

Data were Fourier transformed and deconvolved using the CLEAN algorithm
as implemented in the AIPS routine IMAGR.  The data were weighted using
Brigg's robustness parameter of $-1$, which yields a reasonably small 
point-spread function at the cost of a few per cent loss in sensitivity.
We have also re-analyzed the CHB observations of the region, using the same
data reduction procedures and inputs as were used on the current data.
The results of our analyses are presented in Table \ref{tbl-1} and 
Figures \ref{fig1} \& \ref{fig2}.  
To derive the flux density and position for SN~1961V, a JMFIT
two source Gaussian fit yielded the best results for all 4 observations, while a
single source Gaussian model yielded the best results
 for the other sources in the field of view.  
The positions reported in Table \ref{tbl-1} are weighted averages of the radio 
positions for these sources at the various wavelengths and epochs.
Uncertainties in the peak intensities are reported as the
rms noise from the observations.
To check that changes in measured flux densities are real, we also measured
the flux density of a resolved background source present at
all epochs.  The background source's integrated flux densities for each 
wavelength band are relatively unchanged at both epochs.
Our re-measurements of the CHB data are consistent, within the 
error bars, with those of Cowan \etal (1991).  

\section{Discussion}

We have recovered a radio source at the position of SN~1961V
at 18~cm and 6~cm, coincident, within the error limits, with the CHB position.
Our measured flux densities  
at both wavelengths indicate a clear decline in the 
radio emissions from SN~1961V from the previous CHB observations,
as indicated in Figures \ref{fig1} \& \ref{fig2}.  
The recently measured 18~cm flux, when scaled 
to 20~cm using the newly determined spectral index, indicates a reduction 
in the 20~cm peak flux intensity by 36\% from 1984 to 1999
(see Table \ref{tbl-1}).  The 6~cm
peak flux intensity has also dropped by 54\% in the interval from 1986 to 2000. 
(The western source shows no  change in peak intensity for either
the 6~cm or the 20~cm measurements within noise limits.)
The radio emission from the vicinity of SN~1961V appears to be much more 
complicated than originally thought.  Our new observations and re-analysis of
the CHB data indicate there is at least one previously undetected radio source
within $0\fasec 9$ of SN~1961V.  The radio emission from this source is 
non-thermal at both epochs and has decayed by 50\% at 20~cm and by 33\% at 6~cm.
The region where SN~1961V is located in NGC~1058 is clearly one of recent
star formation.  The peak flux density of this new source near SN~1961V 
is $0.040\pm 0.008$ mJy/beam (at 6~cm) and $0.082\pm 0.026$ mJy/beam 
(at 20~cm).  These values are comparable to that of 
the distinct western source reported by CHB, so this
new source may likely be a previously
undetected SNR.

The decline in the radio flux density of SN~1961V 
is consistent with models for radio emission from SNe (\cite{che84}).  
Synchrotron radiation is produced in the region of
interaction between the ejected supernova shell and the circumstellar 
shell that originated from the prior mass loss of the progenitor star. 
In such models the radio emission drops as the expanding shock wave 
propagates outward through the surrounding and decreasingly dense 
circumstellar material. 
The decline in the flux density of SN~1961V is also consistent with 
Gull's (1973) model for radio emission from SNRs.
This predicts an initial decline in the emission of RSNe for the first 
100 years as the shock overcomes the circumstellar material and a later turn-on
as the build up of material from the ISM results in an increase of synchrotron 
emission, as the object enters the SNR phase.
Thus, the radio emission from these intermediate-age SNe, sources with ages 
comparable to SN~1961V, 
probes the transition region between fading SNe 
and the very youngest SNRs.  In Figure \ref{fig3}, we illustrate 
the radio light curves of several intermediate-aged SNe along with a few SNRs, 
plotting the time since supernova explosion versus the 
luminosity at 20cm.  
It is clear that the radio emission of SN~1961V
at an age of $\simeq$ 38 years is very similar to known
radio SNe at comparable ages, and particularly 
that the radio luminosities of SN~1961V,
in NGC~1058, and the Type II SN~1950B, in M83, are virtually identical at
similar ages.

As shown in Table \ref{tbl-1}, our new observations indicate that SN~1961V  
remains a non-thermal radio source. 
The spectral index, $\alpha$, is relatively unchanged 
although the error bars are rather large.  The spectral index was
derived using the peak intensities, in order to limit the contribution
from the surrounding H II region.
We might expect a possible flattening of the radio spectrum 
as the emission from SN~1961V continues to fade.
This would be an indication of the increasing contribution from the thermal 
emission of the associated H~II region.  Such was the case for the 
radio (\cite{cow94}) and optical (\cite{lon92})
emissions of SN~1957D, in M83, which has now faded below the level of 
an associated H~II region. 
The current and previous values of $\alpha$ for SN~1961V
are still consistent with spectral indices of intermediate-age
RSNe at similar wavelengths, as shown in Table 2.
The non-thermal nature of these sources is well-documented, 
as are those for young radio SNRs, with Cas A, the youngest,  
whose spectral index ranges from $-0.92$ to $-0.64$ 
(\cite{and91}).

We can also compare the rate of decline of radio emission for SN~1961V,
as measured by a  power-law index
($S \propto t^{\beta}$),
with decline rates of known Type II
RSNe (see Figure 3).  The power-law indices for SN~1961V 
were determined from the peak intensities to be 
$\beta _{20cm} =-0.69\pm 0.23$ and $\beta _{6cm} = -1.75\pm 0.16$.
The decay indices for SN~1961V fall within a range 
of previously measured indices for some intermediate-age RSNe (see
Table 2).
In particular, SN~1957D and SN~1970G both have
fairly rapid decline rates, while the younger Type II RSNe 
(SN~1979C and SN~1980K) indicate a slower rate of decline.  We also
note that while the radio emission from SN~1980K has
abruptly dropped after approximately ten years
(\cite{wei92,mon98}),
SN~1979C (at a greater distance than SN~1980K)
is still emitting at detectable levels
(\cite{wei91}).  Recently the radio emission of SN~1980K
appears to have flattened, as indicated in Figure \ref{fig3}.  This may be
a result of the shock wave hitting a denser region of
circumstellar material (\cite{mon00}).
The implications of these comparisons with SN~1961V are that its shock may be 
traveling through considerably more circumstellar material than
similarly-aged RSNe, e.g., SN~1957D.  As a result,  
its radio flux continues to drop at a slower rate more akin
to the younger RSNe, i.e. SNe 1979C, 1980K, and 1986J (the only other
identified Type II$_{pec}$~SN).
Consistent with this interpretation is the very rapid decline in
the radio emissions of Type Ib RSNe, e.g. SN~1983N and and SN~1984L, 
which presumably have less circumstellar material (see Table 2). 
Based on these comparisons the radio observations of SN~1961V are 
consistent with Type II RSNe.
 
Radio comparisons between $\eta$~Car, the super-luminous LBV, and SN~1961V
are more problematic since the first radio observations of $\eta$~Car were 
made 100 years after its eruption.
$\eta$~Car, with a 20 cm flux density of $0.9\pm 0.3$~Jy (\cite{ret83}),
is in fact not a strong radio source when compared to SN~1961V.  
In order to determine $\eta$~Car's 20~cm flux at the current age of SN~1961V,
we have naively assumed a range of potential 
$\beta$ values for $\eta$~Car from $-1$, our measured index for the decline of the flux at 
20~cm of SN~1961V, to $-3$, the index for the decline of SN~1957D.  Applying
these constant decay rates to $\eta$~Car, its 20~cm flux (40 years after
outburst) would range
from 5 to 65 times the Retallack (1983) measurement.  
This would result in $\eta$~Car
being at least 1,000 times weaker than the 
radio source at the position of SN~1961V reported in this paper.
$\eta$~Car's 3~cm flux was measured over a period of 5 years by 
Duncan, White, \& 
Lim (1997) and found to vary between 0.5~Jy and 2.8~Jy, well below the
levels of 6~cm \& 20~cm emissions of SN~1961V.  
They further report 
that $\eta$ Car's spectral index between 3~cm and 6~cm appears 
to peak at $+1.8$ at the position of $\eta$~Car and then drops radially 
toward an index of $0$.  The source of the radio emission is
believed to be thermal radiation from H~II regions associated with $\eta$~Car 
(\cite{ret83}).
The spectral index derived from radio observations at 2~cm and 6~cm of 
Skinner \etal (1998) of P~Cyg, another LBV, is $0.47\pm 0.12$.  (P~Cyg's last
reported outburst was in the 17th century.)
The positive values of the LBV spectral indices are obviously 
very different from 
the negative (i.e. non-thermal) indices for such events as 
SN~1961V, SN~1923A [the 
oldest RSN], and Cas~A [the youngest radio SNR] 
(\cite{and91,cow91,eck98}).  The non-thermal spectral indices for SNe and
SNRs result from a shock front interacting with the CSM and ISM.  As the referee
has pointed out, it is possible that P~Cyg and $\eta$~Car may
have been non-thermal radio sources immediately following their initial
outbursts.  Unfortunately there is no observational evidence to support or
refute this possibility.  Further, it is uncertain whether the radiation 
from an LBV event would remain non-thermal this long after the outburst.
One of the most recent LBV events in the Small Magellenic Cloud, HD~5980, 
was observed in the radio by Ye, Turtle, \& Kennicutt (1991) prior
to LBV outbursts in 1993 and 1994.  It was later observed in 1996 using the
Australian Telescope Compact Array at 3~cm and 6~cm for $\sim$ 1 hour.  
No compact radio emission was detected from the vicinity of the star,
with an upper limit threshold of a few mJys (S. M. White 2001,
private communication).

\section{Conclusions}
Our radio measurements have detected a source at the position of SN~1961V.
The source's radio luminosity, its spectral index, and its decay index are all
consistent with values reported for Type II RSNe and thus appear to support a 
supernova interpretation.  However, the lack of radio observations of similarly-aged bright LBVs prevents a definitive identification of the true nature of SN~1961V.
Additional multiwavelength observations of SN~1961V, as it evolves,
will clearly be needed to make a final judgment about the nature of this 
enigmatic event.  These should include further monitoring with VLA and 
using the Space Telescope Imaging Spectrograph (STIS) to analyze
nebular emission lines from the region near SN~1961V 
to discriminate LBV ejecta nebulae ([N II]-bright),
decades old SNe ([O III] and [O I]-bright), and mature SNRs ([S II]-bright).
The latter observations could be very useful in ruling out one of
the two scenarios for SN~1961V.

\acknowledgments

We thank D. Branch and an anonymous referee for their helpful comments.
The research was supported in part by the 
NSF (AST-9618332 and AST-9986974 to JJC) 
and has made use of the 
NASA/IPAC Extragalactic Database (NED), which is operated 
by the Jet Propulsion Laboratory, 
Caltech, under contract with the National 
Aeronautics and Space Administration.

\begin{deluxetable}{lcc}
\dummytable\label{tbl-1}
\tablenum{1}
\tablewidth{6.5in}
\tablecaption{Radio Observations of the Region Near SN 1961V}
\tablecolumns{3}
\tablehead{
\colhead { }                              &
\colhead{SN~1961V}                        &
\colhead{Western Source}    }
\startdata
Right Ascension (J2000)                        
& $02^{h}43^{m}36\fsec 46 \pm 0\fsec 02$      
& $02^{h}43^{m}36\fsec 24 \pm 0\fsec 01$      \\
Declination (J2000)                            
& $+37\deg 20\amin 43\fasec 2 \pm 0\fasec 2$ 
& $+37\deg 20\amin 43\fasec 8 \pm 0\fasec 4$  \\
\tableline
18 cm\tablenotemark{a} (14 Sept. 1999) 
& $0.147\pm 0.026$               
& $0.117\pm 0.026$               \\
6 cm\tablenotemark{a} (21 \& 25 Jan. 2000)  
& $0.063\pm 0.008$               
& $0.056\pm 0.008$               \\
Spectral Index\tablenotemark{b}   $\alpha ^{18cm} _{6cm}$
& $-0.79\pm 0.23$        
& $-0.64\pm 0.28$                \\
\tableline
20 cm\tablenotemark{a} (15 Nov. 1984)  
& $0.229\pm 0.020$               
& $0.160\pm 0.020$               \\
6 cm\tablenotemark{a} (13 Aug. 1986)  
& $0.135\pm 0.013$               
& $0.070\pm 0.013$               \\
Spectral Index\tablenotemark{b}   $\alpha ^{20cm} _{6cm}$
& $-0.44\pm 0.15$
& $-0.98\pm 0.22$        
\enddata

\tablenotetext{a} {\ peak flux density (mJy/beam)}
\tablenotetext{b} {\ $S \propto \nu^{\alpha}$}

\end{deluxetable}

\clearpage

\begin{deluxetable}{lccc}
\dummytable\label{tbl-2}
\tablenum{2}
\tablewidth{6.5in}
\tablecaption{Comparison with Radio Supernovae}
\tablecolumns{4}
\tablehead{
\colhead {Name}                            &
\colhead{SN Type}                          &
\colhead{Spectral Index ($\alpha$)\tablenotemark{a}}  &
\colhead{Decay Index ($\beta$)\tablenotemark{b}}   }
\startdata
SN~1923A & II$_{P}$   & $-1.00\pm 0.30$ & 
$-6.9\pm 4.0$ (6~cm)            \\
SN~1950B & II?        & $-0.45\pm 0.08$ & 
insufficient observations       \\
SN~1957D & II?        & $-0.30\pm 0.02$ & 
$-2.90\pm 0.07$ (20~cm)         \\
                          &            &                 & 
$-1.70\pm 0.04$ (6~cm)          \\
SN~1961V & II$_{pec}$ & $-0.79\pm 0.23$ & 
$-0.69\pm 0.23$ (20~cm)         \\
                          &            &                 & 
$-1.75\pm 0.16$ (6~cm)          \\
SN~1970G & II         & $-0.56\pm 0.11$ & 
$-1.95\pm 0.17$ (20~cm)         \\
SN~1979C & II$_{L}$   & $-0.72\pm 0.05$ & 
$-0.71\pm 0.08$ (20~cm \& 6~cm) \\
SN~1980K & II$_{L}$   & $-0.50\pm 0.06$ & 
$-0.65\pm 0.10$ (20~cm \& 6~cm) \\
SN~1986J & II$_{pec}$ & $-0.30\pm 0.06$ & 
$-1.18^{+.02}_{-.04}$ (20~cm \& 6~cm) \\
\tableline
SN~1983N & I$_{SL}$   & $-1.0\pm 0.2$ & 
$-1.5\pm 0.3$ (20~cm \& 6~cm) \\
SN~1984L & I$_{SL}$   & $-1.03\pm 0.06$ & 
$-1.59\pm 0.08$ (20~cm \& 6~cm)  \enddata

\tablenotetext{a} {\ $S \propto t^{\beta}$}
\tablenotetext{b} {\ $S \propto \nu^{\alpha}$}
\tablerefs{ \cite{cow91,cow94,eck98,pan86,rup87,wei86,wei92,wei91}}
\end{deluxetable}

\clearpage
\pagestyle{empty}

\clearpage
\begin{figure}
$$ \psboxto(3.1in;0in){fig2a.ps}
\quad \psboxto(0.0in;3.2in){fig2b.ps}$$

\caption{
{\it(a.)} (left) 
Radio contour map at 6 cm (4.9 GHz) of SN~1961V (the strong eastern source) and a neighboring SNR (western source).  Contour levels are -0.07, -0.05, -0.03, 0.03, 0.05, 0.07, 0.10, 0.14, and 0.19 mJy beam$^{-1}$, with a
beam size of $1\fasec 39 \times 1\fasec 17$, p.a. = $-89\deg$, and rms noise of 0.013 mJy beam$^{-1}$.  Observations taken with the VLA in B configuration 13 August 1986.
{\it(b.)} (right) 
Radio contour map at 6 cm (4.9 GHz) of the same region.  Contour
levels are -0.07, -0.05, -0.03, 0.03, 0.05, 0.07, 0.10, 0.14, and 0.19 mJy 
beam$^{-1}$, with a
beam size of $1\fasec 17 \times 1\fasec 13$, p.a.
 = $-35\deg$, and rms noise of 0.008 mJy beam$^{-1}$.  
Observations taken with the VLA in B configuration 21 \& 25 January 2000.
\label{fig2}}
\end{figure}
\clearpage

\clearpage

\begin{figure}
\epsscale{0.9}
\plotone{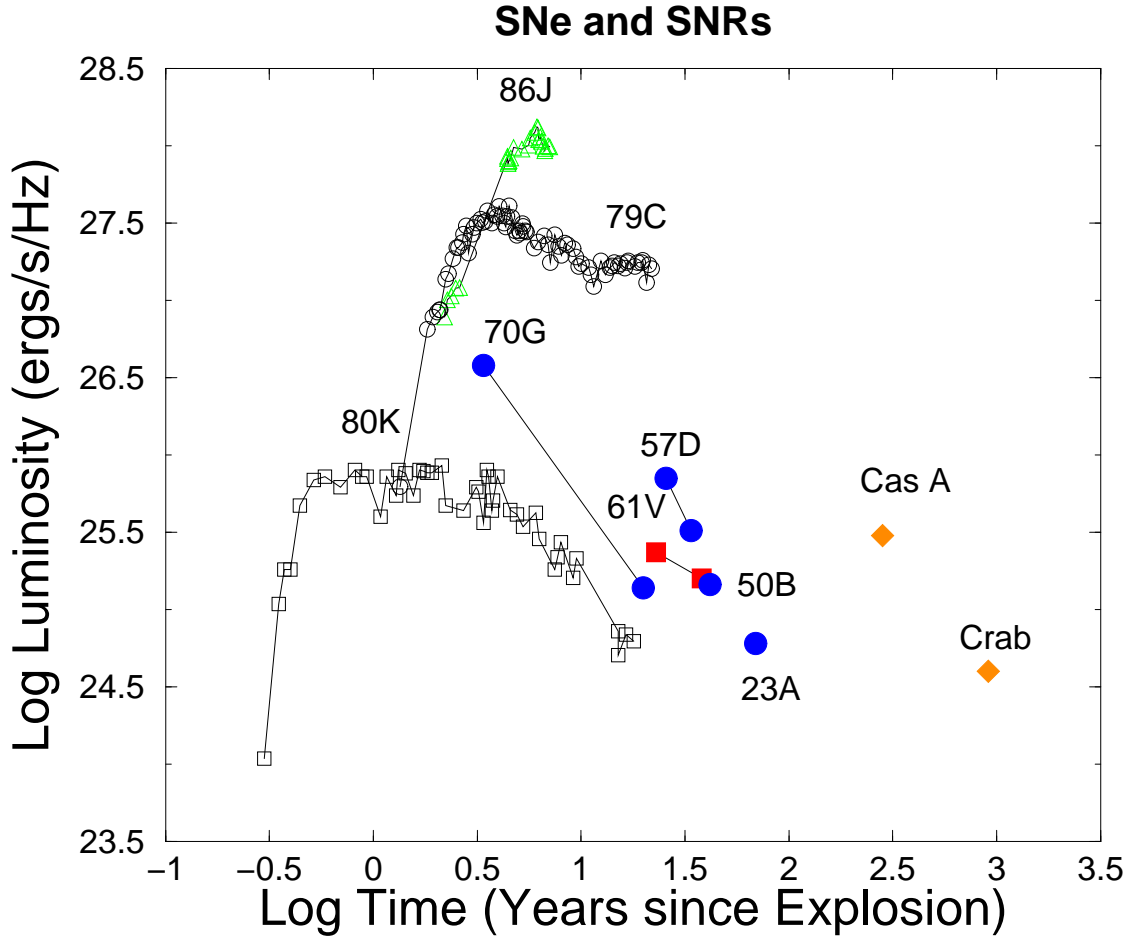}
\caption{
Radio light curve for SN~1961V at 20~cm compared to several RSNe
and SNRs.  Data, fits and distances for SN~1923A, from Eck \etal (1998)
and Saha \etal (1995); for SNe 1950B \& 1957D, from
Cowan \etal (1994) and 
Saha \etal (1995); for SN~1961V, from this paper and 
Silbermann \etal (1996); for SN~1970G, from Cowan \etal (1991) 
and Kelson \etal (1996); for SN~1979C, from 
Weiler \etal (1986, 1991), Montes \etal (2000), and 
Ferrarese \etal (1996); for SN~1980K, 
from Weiler \etal (1986, 1992), Montes \etal (1998), and 
Tully (1988); and for SN~1986J, from Rupen \etal (1987), 
Weiler \etal (1990), and Silbermann \etal (1996).  
Luminosities for Cas~A and the Crab from Eck \etal (1998).
\label{fig3}}
\end{figure}

\end{document}